# End User Computing in AIB Capital Markets: A Management Summary

Andrew McGeady, Joseph McGouran,
AIB Capital Markets,
TSS Architecture & Research,
Riverside IV, 70, Sir John Rogerson's Quay,
Dublin 2, Ireland.
andrew.p.mcgeady@aib.ie
joseph.j.mcgouran@aib.ie

**ABSTRACT**

*This paper is a management summary of how the area of End User Computing (EUC) has been addressed by AIB Capital Markets. The development of an effective policy is described, as well as the process by which a register of critical EUC applications was assembled and how those applications were brought into a controlled environment. A number of findings are included as well as recommendations for others who would seek to run a similar project.*

**Disclaimer: the views expressed in this document are those of the authors and do not necessarily reflect the views of AIB Group or any of its subsidiaries.**

## 1   INTRODUCTION

Traditionally, there was a distinct divide between the roles of *end users* and *professional IT developers* [Grossman, 2002]. However, this line has become blurred over the years, primarily due to the advent of the personal computer and user-friendly development software tools (e.g. Excel, Access, Brio, Crystal Reports, Hyperion) that empower end users (with little or no formal training) to create increasingly complex applications. In many cases these applications are then used as an important information source for key business decisions and internal / external financial reports [Croll, 2005]. This type of application development is generally referred to as "**End User Computing**" (**EUC)** and often takes place in an uncontrolled environment, with poor or non-existent standards for development, testing and change control [Panko & Ordway, 2005]. Some well-publicised cases within the public domain have starkly illustrated the potential impact of poorly controlled EUC [FSA, 2008].

AIB Group [AIB, 2009] is Ireland's leading banking and financial services organisation. Operating principally in Ireland, Britain, Poland and the USA, it employs more than 22,000 people worldwide in more than 750 offices. AIB Capital Markets (AIBCM) comprises AIB Group's Treasury, Investment Banking, Corporate Banking and Asset Management businesses. It is one of four main divisions through which AIB Group operates.

Section 2 of this document offers a short background to the EUC situation in AIB Capital Markets and the creation of an EUC strategy paper. The work of the EUC project in implementing the envisaged strategy is drawn out in sections 3 and 4, including the establishment of an EUC register and the remediation of critical applications. Sections 5 and 6 cover the preparations for future EUC development as well as the project timescale and resources. Section 7 offers a selection of findings and recommendations that may be of use to those with an interest in End User Computing and its application in a real-life scenario.

## 2  BACKGROUND OF EUC IN AIB CAPITAL MARKETS

AIB Group issued an EUC policy in mid-2004. The emphasis of this policy was that if a business unit was to choose to develop mission-critical applications outside the direct management of Divisional IT, the responsibility was put squarely on business management to ensure that such development was carried out in accordance with basic development standards. This policy, while commendable, suffered from a lack of consistency in which controls were demanded and in the way such controls were applied.

In 2006 the AIBCM Architecture and Research team published a strategy document outlining a framework that would provide the necessary control for End User Computing. Five key recommendations were made:

- Acknowledge the EUC issue
- Establish a register of critical EUC applications
- Remediate existing critical EUC applications
- Implement a controlled environment for the housing of such applications
- Develop guidelines for future EUC development

## 3  AIBCM EUC PROJECT

The EUC issue, having come to the fore after the introduction of compliance legislation such as SOX and the heightened vigilance of auditors [Alliy & Brown, 2008] in this area, was acknowledged by AIBCM as a key business objective. The strategy paper published by Architecture & Research was adopted as a model to follow within AIBCM and led to the establishment of the EUC Project. The project's objectives were based on the key recommendations of the strategy document.

### 3.1  Establishing the EUC Register

The EUC policy required that there be a register of critical EUC applications established for Capital Markets. In order to lessen the burden on business units it was decided to employ an outside consultancy to facilitate this exercise. The register was the result of the following process:

**Communications**

The EUC project team visited each business unit in the company, carrying out briefings at various levels from management board down to ground level. The purpose of such briefings was two-fold:

1) to create an awareness of the EUC policy and its implications
2) to acquire a list of appropriate EUC contacts for each unit, people with whom the project could work to surface critical EUC applications

**Data Gathering**

When EUC contacts were made available for each business unit, the project team further briefed each contact to ensure that they understood the precise task that was their charge. EUC contacts were sent a copy of the EUC policy, a copy of a risk-based criticality assessment as well as an application information template.

Contacts were required to enter eighteen pieces of data per application into the template. Each entry in this template would ultimately form the basis of an entry in the EUC register.

**Workshops**

Upon receipt of a completed application information template, a workshop was scheduled with the EUC contact, a member of the project team and an external EUC consultant in attendance. During a workshop applications were discussed in more detail in order to gain more subtle information about them. The workshop process gave the business contacts an opportunity to further refine their lists of critical applications – the project team could not determine criticality on behalf of the business however such a setting was conducive to the seeking and supplying of guidance in such matters, as well as discussion of the risk assessment.

The assessment was used to aid the contact in what might and might not be deemed critical. This was in recognition of the fact that even though one might use an application in the course of one's daily tasks, that does not make such an application, in and of itself, mission critical to the business. In some areas this was recognised immediately by the EUC contact during the workshop whereas in others the call was made later at Executive level.

*Scale*

While industry sensitivities understandably restrain the authors from providing precise figures, that which can be stated is that in AIB Capital Markets there exist approximately five times the number of critical EUC applications as those that would be classified as "Sox Tier-1" applications. This ratio is not consistent, however, as it varies between business units and departments.

**Initial Remediation Plan**

As well as providing guidance and collecting further information about an application, the job of the EUC consultant was to produce an initial remediation plan for each application. Categories into which an application might be placed were as follows:

- *Document:* the application requires procedural and/or technical documentation.

- *Test:* the application requires testing to ensure that it performs the stated function.

- *Control:* the application will be migrated to a controlled IT environment.

- *Minor Enhancement:* the application requires minor enhancements without the need for significant involvement from business users. Examples would include providing basic input validation control, locking cells once the report is prepared, etc.

- *Enhance:* the application requires enhancements to its functionality. These would include building interfaces to other systems, automating report generation, etc.

- *Migrate:* the application should be migrated to a different platform as the current platform cannot support the required features (e.g. Access -> SQL Server).

- *Replace/New Development:* the platform is unsuitable for the required task. The application should be replaced either by extending the functionality of an existing IT-supported system or by a new development on a more robust platform.

**Validation**

Each workshop report, having been prepared by the EUC consultant, was sent back to the relevant EUC contact so that they might verify that there were no mistakes, misunderstandings or misinterpretations made by the project team. Upon validation each report was uploaded to the EUC register.

**Controlled Environment**

It was recognised that, in order to best assist end users in controlling critical EUC applications, a technical solution would be required. Automation was the key, with the various controls being viewed as useful aids or tools rather than as bureaucratic burdens to be evaded wherever possible. AIB Capital Markets have implemented an automated spreadsheet auditing solution and are exploring the possibility of the further integration of EUC controls, such as version control and access control, with the wider file environment. This solution is referred to as the EUC controlled environment.

**4   REMEDIATION**

With an EUC register having been put in place, as well as an EUC controlled environment, the remaining task was the remediation of existing critical EUC applications. At its most basic level, this would involve taking each application on the AIBCM EUC Register and putting it through a process of documentation [Payette, 2006] and testing [Pryor, 2003, 2004] before migration to the controlled environment.

It was recognised at an early stage in the project that there would be resourcing issues for such remediation work, both from the point of view of IT and the Business. Therefore it was decided to engage with an external EUC consultancy to partner in this effort.

A remediation process was developed that would have minimal impact on the day-to-day business of the company. Briefly summarised:

- A copy of each application was sent to the project team.

- Each application was analysed, both from a technical and functional perspectives, in consultation with the application owner.

- The resulting documentation was broken down into three parts:
    - Functional Specification
    - Technical Specification
    - User Guide

- Where "applications" comprised multiple spreadsheets, a single documentation set was created for the sake of simplicity.

- The only impacts on the business were the initial consultation, validation of documentation and test results as well as any necessary follow-up.

Each application on the critical register was brought through the remediation process, with each business area charged back directly per unique application for the effort involved.

## 5 CONTROLLING FUTURE EUC DEVELOPMENT

### 5.1 Guidelines

Development of critical IT applications should be carried out by a central IT function. However, in certain areas of an organisation it may be appropriate that business users with expert domain knowledge would be permitted to develop critical applications, provided such users are given sufficient IT development training and development is carried out in a controlled manner.

The development of a critical application, whether by a central IT department or by a junior clerk in the front office, must adhere to certain basic principles. In order to better control future EUC development it was necessary to design a framework that would cater for such principles while not obstructing the day-to-day work of the business. AIB Capital Markets have designed a version of the Software Development Life Cycle (SDLC) that has been made appropriate for EUC, spreadsheets in particular [Grossman, 2002]. Guidelines and templates, consistent with the EUC policy and the central tenets of the SDLC, will be put into a handbook, an 'all you need to know about EUC' resource, and made available to all on the company intranet upon ratification of the 2008 EUC policy.

### 5.2 EUC Development Community

By making EUC development both credible and controllable, and increasing the expertise of business EUC developers, new working relationships have been created. This EUC development community will be nurtured, as there may well be future projects for which a relationship with business "power users", harnessing their unique mix of business knowledge and IT expertise, would be hugely beneficial for the organisation.

## 6 Timescale & Resources

EUC project work in AIB Capital Markets began in 2006, with all external resources withdrawn as of December 2007. AIBCM committed one staff member to the project full-time, as well as two staff members on a part-time basis. The external consultancy provided a project manager as well as six analysts, both business and technical.
The EUC Co-Ordinator role remains, whose job it is to maintain the EUC register, promote the EUC policy and provide consultation and advise regarding any EUC issues within the division.

## 7 SUMMARY OF FINDINGS

### 7.1 Application Ownership

It is important that there should be no confusion regarding application ownership. Business owners do not want to cede control over *their* applications while IT does not want to become a foster parent to applications in whose development it has played no previous part. There is also the question of who can properly judge the risk of allowing such development to proceed.

In the case of AIBCM these divisions have been made clear in a new policy to be ratified in 2008:

- the integrity of an end-user-developed application will always remain the ultimate responsibility of the business owner of that application.

- Operational Risk is now responsible for permitting a business area to develop critical EUC applications. This permission is granted based on certain criteria being fulfilled, for example the level of technical expertise in the department and the suitability of the proposed critical applications for development outside a central IT function.

- where IT provides a technical framework to better assist in the control of such applications, IT is responsible for the integrity of that framework while the Business remains responsible for each application contained within.

The key to avoiding confusion is to ensure that divisions of ownership and responsibility are both straightforward and logical and set out clearly in an organisation's EUC policy.

**7.2     Message**

When discussing EUC it is necessary to make clear that the intention is to introduce a practical, appropriate level of control. Although similar in principle EUC is not the same as SOX (Sarbanes-Oxley Act of 2002). EUC is broader than SOX, an inexact superset to be precise, however SOX is far deeper in its requirements. The effort expended in becoming 'SOX-compliant' has placed a massive strain on companies, departments and individuals worldwide. Not to put too fine a point on it, the word 'SOX' can mean death to EUC from the point of view of selling the benefits of EUC controls to business users.

**7.3     The Personal Touch is Vital**

Making the effort to talk personally to all involved is of vital importance in ensuring the success of EUC in the organisation. This is, of course, true for most projects however End User Computing is a concept filled with subtleties that are not often best expressed through characters either on screen or in print. The effort spent personally briefing each management board, executive, IT manager and EUC contact goes to ensure that they are in a position to assist the project, they are enthusiastic about (or at the very least not opposed to) the notion of introducing controls for EUC and that there will be a good standard of information returned to the project, reducing the time required to return to the business for yet more consultation.

**7.4     Program Sponsor**

The introduction of EUC controls must be recognised as being a business solution to a very real business issue. In order that EUC controls are not perceived by the business to be a case of IT sticking its nose in where it's not wanted, EUC control initiatives must be seen to be driven by business brokers, be they from finance, risk or some other area. This bestows business credibility to such initiatives, as well as to those who might facilitate their implementation.

## 8 CONCLUSION

The goal of any organisation with regard to the development of critical IT applications should be that such development is carried out by trained IT developers. However, the development of critical applications by end users is an issue that must be acknowledged by organisations. End User Computing has arrived; it is here to stay. It can either be performed in a controlled manner, serving to advance organisational goals, or performed "in the dark", serving only to add to the level of risk carried by the firm. It is the view of the authors of this paper that in certain circumstances, with appropriate controls in place, end user development can be a valuable asset to the organisation, combining in-depth business knowledge with the power of IT to create a development model that can happily complement the existing traditional IT development processes.